\documentclass[twoside]{photon2007}
\usepackage[latin1]{inputenc}
\usepackage[dvips]{graphicx,epsfig,color}
\usepackage{wrapfig,rotating}
\usepackage{amssymb,amsmath,array}

\makeatletter
    \def\thebibliography#1{\section*{Bibliography\@mkboth
      {REFERENCES}{REFERENCES}}\list
      {[\arabic{enumi}]}{\settowidth\labelwidth{[#1]}\leftmargin\labelwidth
	\advance\leftmargin\labelsep
	\usecounter{enumi}}
	\def\newblock{\hskip .11em plus .33em minus .07em}
	\sloppy\clubpenalty4000\widowpenalty4000
	\sfcode`\.=1000\relax}
    \makeatother

\begin{document}
\title{
Photoproduction in Ultra-Peripheral Relativistic Heavy Ion Collisions with STAR}
\author{Janet Seger (for the STAR Collaboration)
\vspace{.3cm}\\
Creighton University \\
Omaha, NE, USA
}
\maketitle

\begin{abstract}
We present a summary of recent photoproduction results in ultra peripheral relativistic heavy ions collisions with STAR.  These collisions have impact parameters larger then twice the nuclear radius; the nuclei do not physically collide, but interact via long-range electromagnetic fields.  We observe exclusive $\rho^0$ production as well as $AuAu\rightarrow Au^*Au^* \rho^0$ with accompanying mutual nuclear excitation at $\sqrt{s_{NN}}=200$ GeV.  We report the $\rho^0$ production cross section for both coherent and incoherent coupling accompanied by mutual nuclear excitation.  We have studied the cross section as a function of $p_T$, $y_{\rho^0}$ and $M_{\pi\pi}$ and compared it to theoretical models. In addition, we measured the $\rho^0$ helicity matrix elements.  They are found to be consistent with s-channel helicity conservation.  The ratio of coherent $\rho^0$ and direct $\pi^+\pi^-$ pair photoproduction has been measured and found to be consistent with earlier measurements.  The 4-pion final state $AuAu \rightarrow \pi^+\pi^-\pi^+\pi^-$ state has also been observed.
\end{abstract}

\section{Introduction}
The Relativistic Heavy Ion Collider (RHIC) collides completely ionized heavy nuclei at relativistic speeds.  A variety of systems and energies have been studied:  Au-Au at  $\sqrt{s_{NN}}=$ 19.6, 62.4, 130 and 200 GeV, Cu-Cu at $\sqrt{s_{NN}} =$ 62.4 and 200 GeV and d-Au at $\sqrt{s_{NN}} =$ 200 GeV.  In addition, polarized protons have been studied at $\sqrt{s_{NN}}=$ 200 and 410 GeV.  I will focus on data collected with the Solenoidal Tracker At RHIC (STAR) detector from Au-Au collisions at 200 GeV.

The STAR detector includes a Time Projection Chamber~\cite{tpcdes} that records charged tracks in a cylindrical volume with a 2 m radius and a length of 4 m.  A Central Trigger Barrel~\cite{trigdes} surrounds the Time Projection Chamber with 240 scintillator strips.  Two hadronic calorimeters are placed near the beamline at $\pm 18$ m from the interaction point.  These ``zero-degree calorimeters" (ZDCs)~\cite{zdcdes} detect neutrons from the collision.

An ultra-peripheral collision occurs when the two colliding nuclei pass in close proximity but do not geometrically overlap.  Very intense electric fields act for a short period of time, exchanging virtual photons that can be described in the Weizs\"{a}cker-Williams formalism~\cite{ww}.  The photons are very nearly real.  This photon exchange can result in lepton or meson pair production, vector meson production and/or Coulomb excitation of one or both nuclei.  Typically, these events result in at most a few charged tracks in the Time Projection Chamber.

\begin{figure}[htb]
\centerline{\includegraphics[width=0.7\columnwidth,totalheight=0.15\textheight]{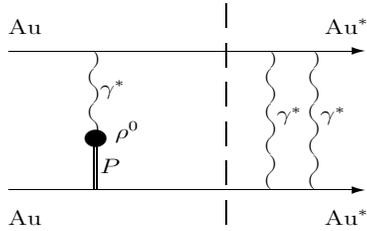}}
\caption{Feynman diagram of photonuclear reaction producing a $\rho^0$-meson.  Independently exchanged photons result in nuclear excitation leading to neutron emission.}\label{fig:rhoF}
\end{figure}

Vector mesons can be produced through a photonuclear interaction~\cite{baur, bertkn}.  In this case, one nucleus emits a photon, which fluctuates to a virtual $q\bar{q}$ pair.  This pair then scatters from the other nucleus, producing a real vector meson.  For coherent production, a limit is placed on the transverse momentum, $p_T < \hbar/2R_A \sim 150$ MeV.  This condition allows coherently produced vector mesons to be separated experimentally.  Additional exchanged photons may also leave the nuclei in excited states that decay by neutron emission~\cite{nbk,multiphoton}.  This is shown in Fig. \ref{fig:rhoF}.  The exchange of two photons resulting in the production of $e^+e^-$ pairs has been studied at 130 GeV~\cite{epem}; this process is a negligible background to $\rho^0$ production at 200 GeV.
The cross section for these interactions factorizes and can be written as
\begin{eqnarray}\label{eq:crossprob}
\lefteqn{\sigma(\text{Au Au} \rightarrow \text{Au}^{*}\text{Au}^{*} \rho^{0})= } \nonumber\\ & = \int d^{2}b
\big[1- P_{\text{Had}}(b)\big] P_{\rho^{0}}(b)P_{exc}(b)
\end{eqnarray}
where   $P_{\text{Had}}(b)$   is  the   probability   of  a   hadronic
interaction,   $P_{\rho^{0}}(b)$  is  the   probability  to   produce  a
$\rho^{0}$, and $P_{exc}(b)$  is the probability of nuclear excitation.

STAR uses several triggers to study ultra-peripheral collisions.  Our ``topology trigger"~\cite{130gev}, requires a low total multiplicity combined with a coincidence in North and South quadrants of the Central Trigger Barrel.  Events with coincidences in the top and bottom quadrants are vetoed to avoid contamination from cosmic rays.

Our other triggers rely at least in part on a coincidence in the ZDCs, and are therefore sensitive only to those processes that produce nuclear excitation.  Our ``minimum bias trigger" combines a small total multiplicity in the Central Trigger Barrel with a coincidence in the ZDCs.  Our ``multi-prong trigger" combines coincident neutrons in the ZDCs, a low multiplicity in the Central Trigger Barrel, and a veto in the Beam-Beam Counters.

\section{Coherent production of $\rho^0$ mesons}
To select $\rho^0$ production events, we require 2 tracks with opposite charge that are back-to-back in the transverse plane, originate from a common vertex and have a low total transverse momentum.  The primary backgrounds are cosmic rays (which can be reduced with a ZDC requirement, or a cut around y=0), beam-gas interactions (which can be reduced with multiplicity and vertex cuts) and hadronic interactions (which can be reduced with cuts on multiplicity and $p_T$).  The $\rho^0$ mass is fit with the function

\begin{figure}[htb]
\begin{picture}(200,110)
\centerline{\includegraphics[width=0.8\columnwidth,totalheight=0.2\textheight]{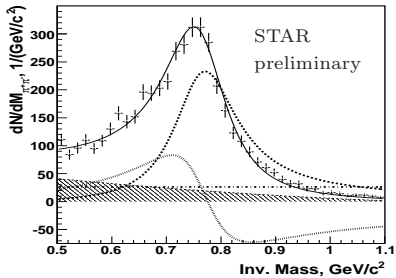}}
\put(-80,90){\scriptsize STAR }
\put(-80,80){\scriptsize preliminary}
\normalsize
\end{picture}
\caption{Invariant mass distribution for the minimum bias dataset for 200 GeV Au-Au collisions.  The points are data (error bars are statistical only).  The dashed line shows the Breit-Wigner contribution, the dot-dashed line shows the contribution from direct pion production, while the dotted line shows the interference from direct pion production.  The gray histogram shows the background contribution.  The solid line shows the best fit to the data.}\label{fig:massfit}
\end{figure}

\begin{scriptsize}
\begin{equation}\label{eq:fitfunc}
\frac{dN}{dM_{\pi\pi}}=\left| A\frac{\sqrt{M_{\pi\pi}M_{\rho^{0}}\Gamma_{\rho^{0}}}}{M_{\pi\pi}^{2}-M_{\rho^{0}}^{2}+iM_{\rho^{0}}\Gamma_{\rho^{0}}}+B \right|^{2} + f_{bg}
\end{equation}
\end{scriptsize}
This includes a Breit-Wigner function~\cite{ZEUSspindensity} for the signal and a S\"{o}ding interference term~\cite{soding, sakurai} to account for direct $\pi^+\pi^-$ production.  Here $A$ is the amplitude for resonant $\rho^0$ production, $B$ the amplitude for direct pion pair production and $\Gamma_{\rho^0}$ is the momentum dependent width of the $\rho^0$.  A second order polynomial, $f_{bg}$, is used to describe the background, which has been estimated with like-sign pairs.  There are approximately 16,000 candidates in the topology and minimum bias samples combined.  Fig.  \ref{fig:massfit} shows the fit to the invariant mass distribution for the minimum bias dataset taken in the 2001 200 GeV Au-Au run.
The ratio of direct pion production to resonant $\rho^0$ production is measured experimentally through this fit.  In 200 GeV Au-Au collisions, STAR measures $|B/A| = 0.84 ± 0.11 \text{ GeV}^{-1/2}$ (STAR preliminary) and at 130 GeV, $|B/A| = 0.81 \pm 0.28 \text{ GeV}^{-1/2}$~\cite{130gev}.  These results are in agreement with each other, showing no energy dependence in this ratio.  There is also no angular dependence or rapidity dependence, in agreement with ZEUS measurements~\cite{ZEUSspindensity}.  Coherent $\rho^0$ production has also been studied in $d-Au$ collisions~\cite{timosh}.

There are three published model predictions for the $\rho^0$ cross section.  The Klein and Nystrand (KN) model~\cite{ksjn} is a vector dominance model with a classical mechanical approach for scattering, based on $\gamma p\rightarrow \rho p$ experiments results.  The Frankfurt, Strikman, Zhalov (FSZ) model~\cite{fsz} is a generalized vector dominance model with a Gribov-Glauber approach.  The Gon\c{c}alves and Machado (GM) model~\cite{gm} includes a QCD dipole approach with nuclear effects and parton saturation phenomenon.  The three models predict significantly different values for the total production cross section, as well as different rapidity distributions in the total cross section.

\begin{figure}[htb]
\begin{picture}(200,110)
\put(0,0){\includegraphics{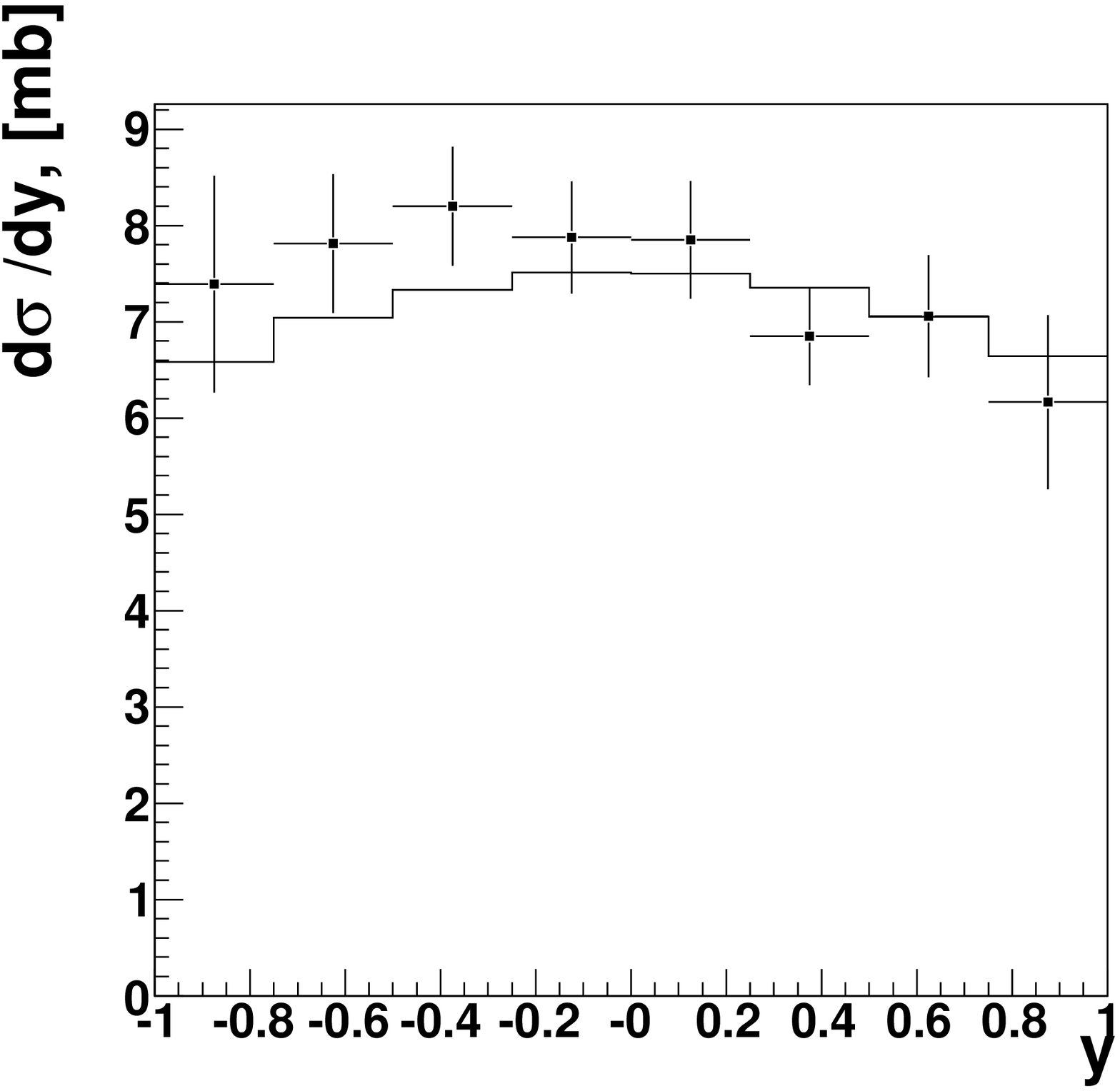}}
\put(95,0){\includegraphics{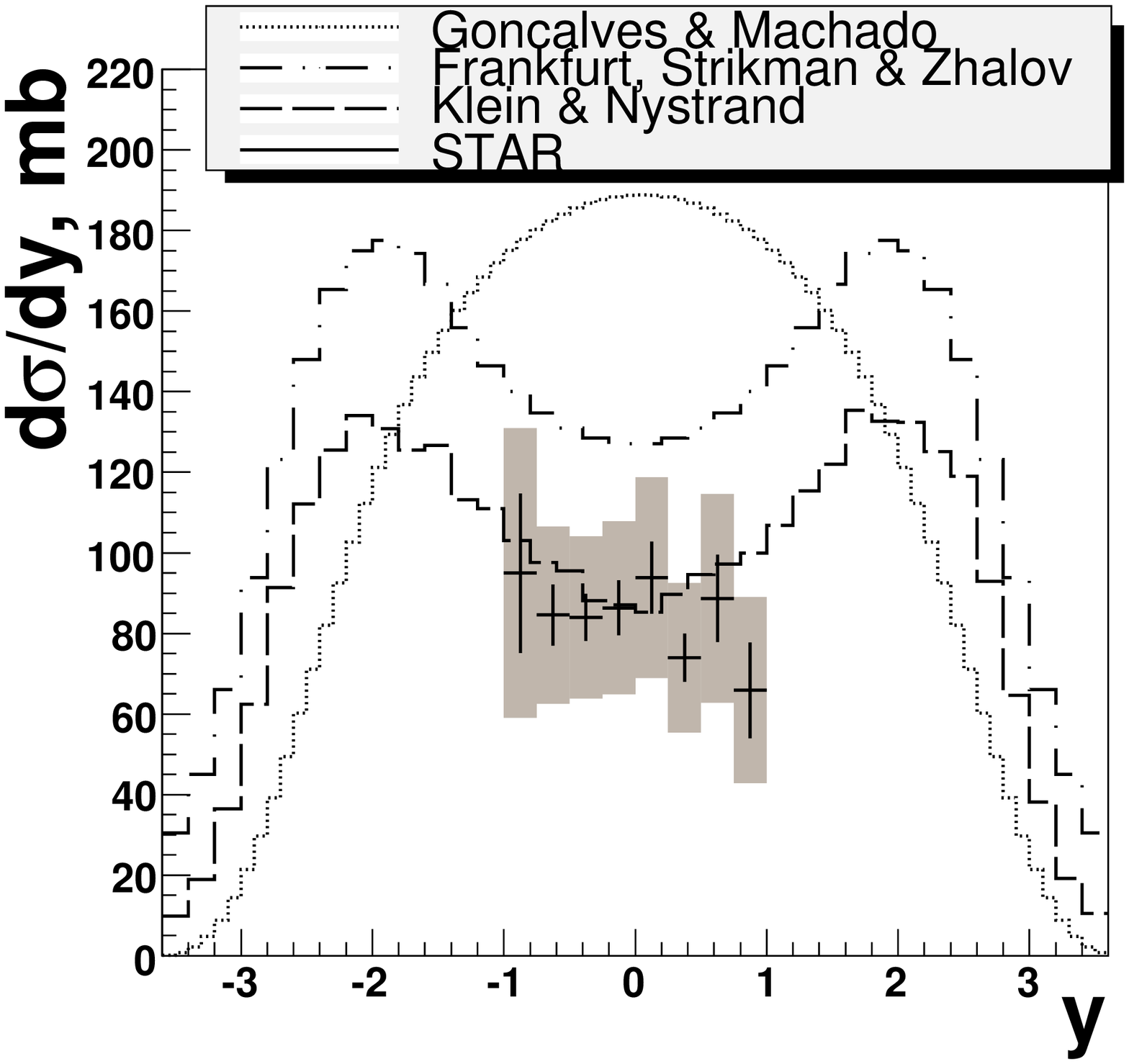}}
\put(125,20){\tiny STAR preliminary}
\put(37,33){\tiny STAR}
\put(33,23){\tiny preliminary}
\normalsize\end{picture}
\caption{The left panel shows the measured rapidity distribution (points) for events with mutual Coulomb excitation along with a Monte Carlo based on the KN model (solid line).  The right panel shows the corresponding distribution for events without requiring that Coulomb excitation occur.  Error bars are statistical only. The gray shaded areas on the right panel indicate the systematic errors. The three curves show the theoretical predictions. }\label{fig:rapidity}
\end{figure}

Fig. \ref{fig:rapidity} shows the rapidity distribution for $\rho^0$ mesons accompanied by mutual Coulomb excitation in the experimentally measured region ($|y| < 1$).  Superimposed on the data is a Monte Carlo based on the KN model.  To compare our data to the theoretical predictions, it is necessary to extend this measurement in two ways:  1) to the total cross section, by including events with no Coulomb excitation or excitation of only a single nucleus, and 2) to full rapidity.

In our ``topology trigger" data, which places no restriction on the signal in the ZDCs, we can measure the relative cross sections for the different excitation states as a function of rapidity.  It is difficult to measure an absolute cross section in this data sample because the trigger efficiency is poorly known.  However, we can use this ratio to scale the mutual-excitation cross section up to the full cross section.  The right panel in Fig. \ref{fig:rapidity} shows the total cross section in the experimentally-measured region, with predictions from the three models superimposed.  Unfortunately, within the measured rapidity region the experimental data cannot distinguish between the models on the basis of the shape of the distribution.

The predicted values for production cross sections include full rapidity.  Scaling from our measured rapidity region to full rapidity is necessarily model dependent.  We use an extrapolation factor derived from the shape of the rapidity distribution in the KN  model.  The FSZ model predicts a very similar shape and would result in an extrapolation factor that differs by only $3\%$ from that derived from the KN model.
The cross section for $\rho^0$ meson production accompanied by mutual excitation in Au-Au collisions at 200 GeV in the experimentally measured region of $|y| < 1$ is $30.3 \pm 1.1 \pm 6.4$ mb (STAR preliminary), to be compared with the 130 GeV cross section of $26.2 \pm 1.8 \pm 5.8$ mb~\cite{130gev}.  The total cross section, extrapolated to full rapidity is $509 \pm 35 \pm 107$ mb at 200 GeV (STAR preliminary), as compared to $410 \pm 190 \pm 100$ mb at 130 GeV~\cite{130gev}.  The KN model predicts a total cross section with full rapidity coverage at 200 GeV of 590 mb~\cite{ksjn}, consistent with our measured results.  The FSZ model predicts a somewhat higher cross section of 876 mb~\cite{fsz}, and the GM model predicts an even higher cross section of 934 mb~\cite{gm}.

\begin{figure}[htb]
\begin{picture}(200,90)
\put(0,0){\includegraphics{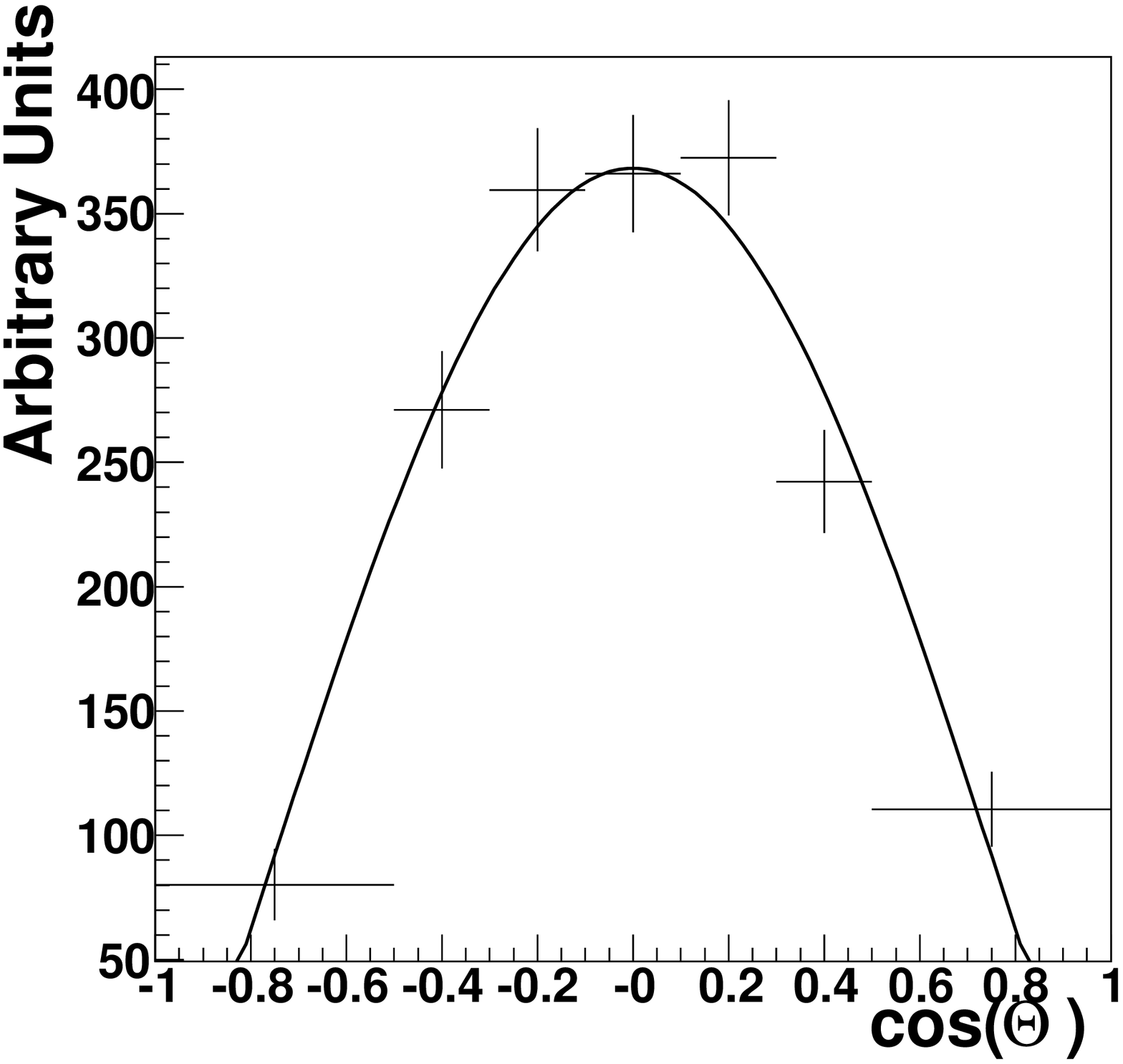}}
\put(95,0){\includegraphics{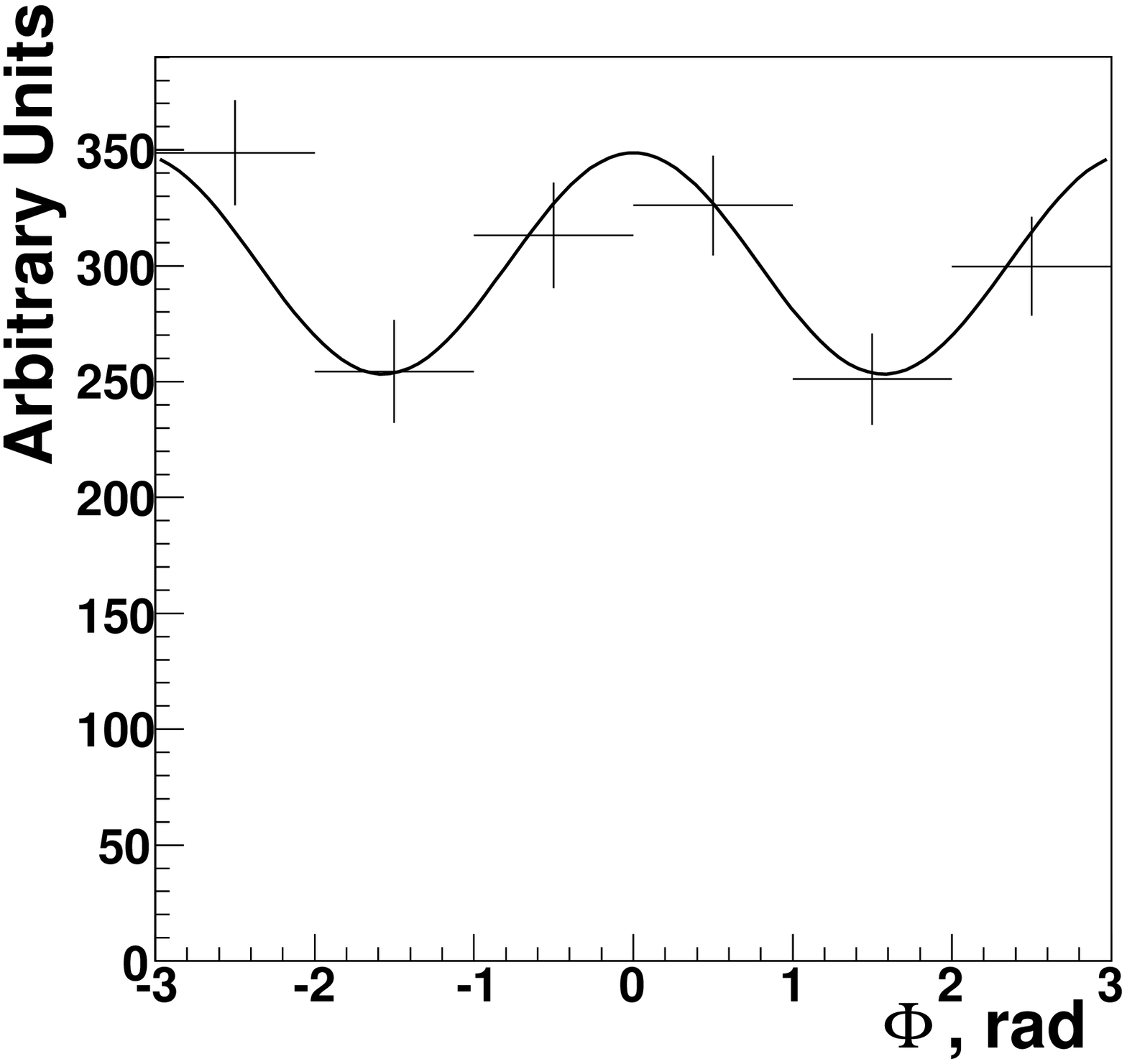}}
\put(115,33){\scriptsize STAR preliminary}
\put(43,33){\tiny STAR}
\put(39,23){\tiny preliminary}
\normalsize
\end{picture}
\caption{The left panel shows the projection of the differential cross section onto $cos(\Theta_h)$.  The right panel shows the projection on to $\Phi_h$.  The points are the data; error bars are statistical only.  The smooth curve shows the fit.}\label{fig:spindensity1}
\end{figure}

We have also measured the $\rho^0$ meson spin density matrix elements.  These offer a way to test whether or not s-channel helicity is conserved and the vector meson therefore retains the helicity of the exchanged photon~\cite{gilman}.  By measuring the decay angular distribution in the rest frame of the $\rho^0$ meson, we are able to determine the 3 (of 15) independent spin density matrix elements.  The matrix elements are measured by fitting the projections of the differential cross sections with Equation \ref{eq:fitfunchelicity}~\cite{schil}
\begin{eqnarray}\label{eq:fitfunchelicity}
 \frac{1}{\sigma}\frac{d^2\sigma}{d \cos(\Theta_{h}) d \Phi_{h}}=\frac{3}{4\pi} \cdot [\frac{1}{2}(1-r^{04}_{00}) \nonumber \\
+\frac{1}{2}(3r^{04}_{00}-1)\cos^{2}(\Theta_{h}) \nonumber \\
-\sqrt{2}\Re e[r^{04}_{10}]\sin(2\Theta_{h})\cos(\Phi_{h}) \nonumber \\
-r^{04}_{1-1}\sin^{2}(\Theta_{h})\cos(2\Phi_{h})].
\end{eqnarray}
where $\Theta_h$ is the polar angle between the ion and direction of the $\pi^+$, $\Phi_h$ is the azimuthal angle between decay plane and production plane, $r_{00}^{04}$ represents probability the $\rho^0$ has helicity 0, $r_{1-1}^{04}$ is related to the level of  interference between helicity non-flip and double flip, and $\Re e[r_{10}^{04}]$ is related to the level of  interference between helicity non-flip and single flip.  The fits are shown in Fig. \ref{fig:spindensity1}.
If s-channel helicity conservation holds, all 3 of these measured spin-density matrix elements are expected to be approximately zero.  The STAR preliminary values are given in Table \ref{tab:spinelem}.  They are all consistent with zero, indicating that s-channel helicity conservation holds.
\begin{table}[h]
\begin{scriptsize}
\begin{tabular*}{200 pt}{lcc}
Param. & Fit result & $\gamma p $ experiment \\ \hline
$r^{04}_{00}$      & -0.03 $\pm$ 0.03 $\pm$ 0.06  & 0.01 $\pm$ 0.01 $\pm$ 0.02    \\
$\Re e[r^{04}_{10}]$ & 0.04 $\pm$ 0.02 $\pm$ 0.03  & 0.01 $\pm$ 0.01 $\pm$ 0.01  \\
$r^{04}_{1-1}$     & -0.01 $\pm$ 0.03 $\pm$ 0.05 & -0.01 $\pm$ 0.01  $\pm$ 0.01\\
\end{tabular*}
\end{scriptsize}
\caption{\label{tab:spinelem}STAR preliminary values for 3 spin density matrix elements for $\rho^0$ production in 200 GeV Au-Au collisions along with comparison values for $\rho^0$ production in $\gamma p $ collisions measured by ZEUS~\cite{ZEUSspindensity}. The first error is statistical, the second is systematic.}\end{table}

\section{Incoherent production of $\rho^0$ mesons}
	We can investigate incoherent production of $\rho^0$ mesons in ultra-peripheral collisions by extending the range in transverse momentum considered for analysis.  Below $p_T$ of about 150 MeV, most of the production is coherent.  By extending the analyzed range to approximately 550 MeV, we can compare coherent and incoherent production.  We fit the $p_T^2$ distribution to a double exponential function
\begin{equation}
{d^{2}\sigma\over {dydt}} = A_{\text{c}}\exp{(-B_{\text{c}}t)} + A_{\text{i}}\exp{(-B_{\text{i}}t)}.
\label{eq:dsdt}
\end{equation}
\begin{figure}[htb ]
\begin{picture}(200,110)
\centerline{\includegraphics[width=0.85\columnwidth,totalheight=0.25\textheight]{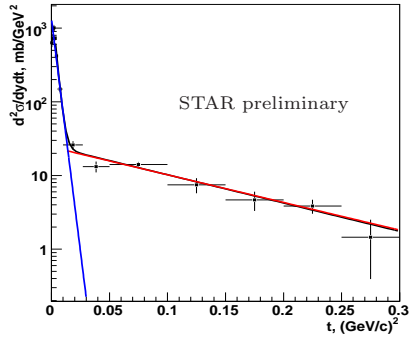}}
\put(-110,90){\scriptsize STAR preliminary}
\normalsize
\end{picture}\caption{The $t_{\perp} =p_T^2$ distribution for $\rho^0$ production accompanied by mutual Coulomb excitation in Au-Au collisions at 200 GeV.  The data is fit to a double exponential.  The steep line at low $t$ shows the contribution from coherent production; the more gradual line at higher $t$ shows the contribution from incoherent production.}\label{fig:incoherent}
\end{figure}

This is shown in Fig. \ref{fig:incoherent}.  The fit parameter $B_{\text{c}}$ describes the exponential that dominates at low-$p_T$, corresponding to coherent production.  The fit parameter $B_{\text{i}}$ describes the exponential that dominates at high $p_T$, corresponding to incoherent production.  We measure $B_{\text{i}} = 8.8 \pm 1.0 \text{ GeV}^{-2}$ and $B_{\text{c}} = 388.4 \pm 24.8 \text{ GeV}^{-2}$ and the ratio of incoherent to coherent production $\sigma(\text{incoh})/\sigma(\text{coh}) ~ 0.29 \pm 0.03$ (STAR preliminary).  The fitted region does not extend below $t=0.002 GeV^2/c^2$, since the interference  between  $\rho^{0}$ photoproduction  from  the two  nuclei reduces  $d^{2}\sigma/dydt$ at  small  $t$~\cite{KNpt,  nbk}.

\section{Coherent Production of $\pi^+\pi^-\pi^+\pi^-$ }
	In addition to $\rho^0 \rightarrow \pi^+\pi^-$, STAR has also observed the process $\gamma Au \rightarrow  \pi^+\pi^-\pi^+\pi^-$.  This data was collected with our ``multi-prong" trigger.  The signature for coherent production is 4 charged tracks reconstructed in the Time Projection Chamber that have a total charge of zero and a total $p_T < 150$ MeV/c.

\begin{figure}[htb]
\begin{picture}(200,90)
\put(0,0){\includegraphics{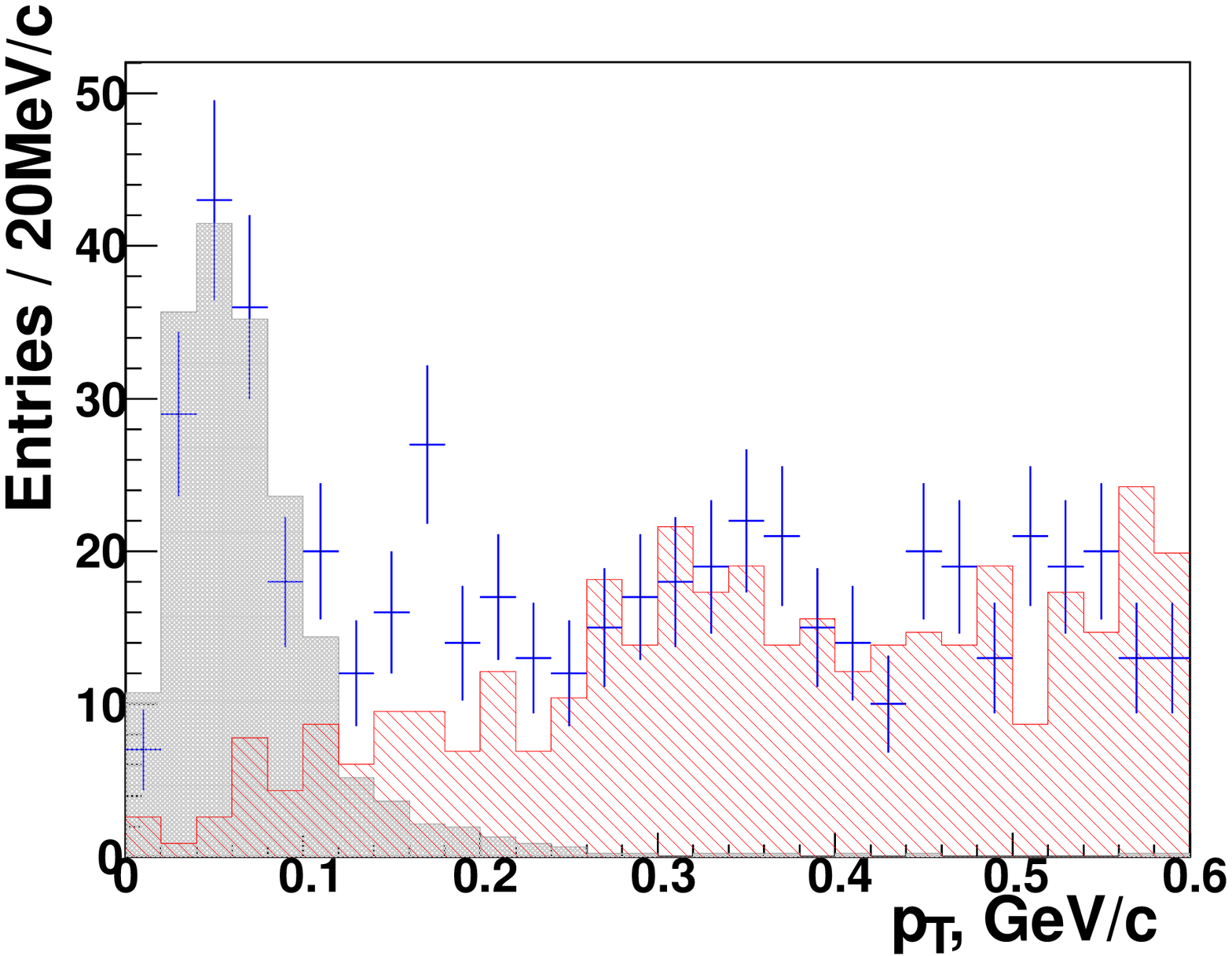}}
\put(95,0){\includegraphics{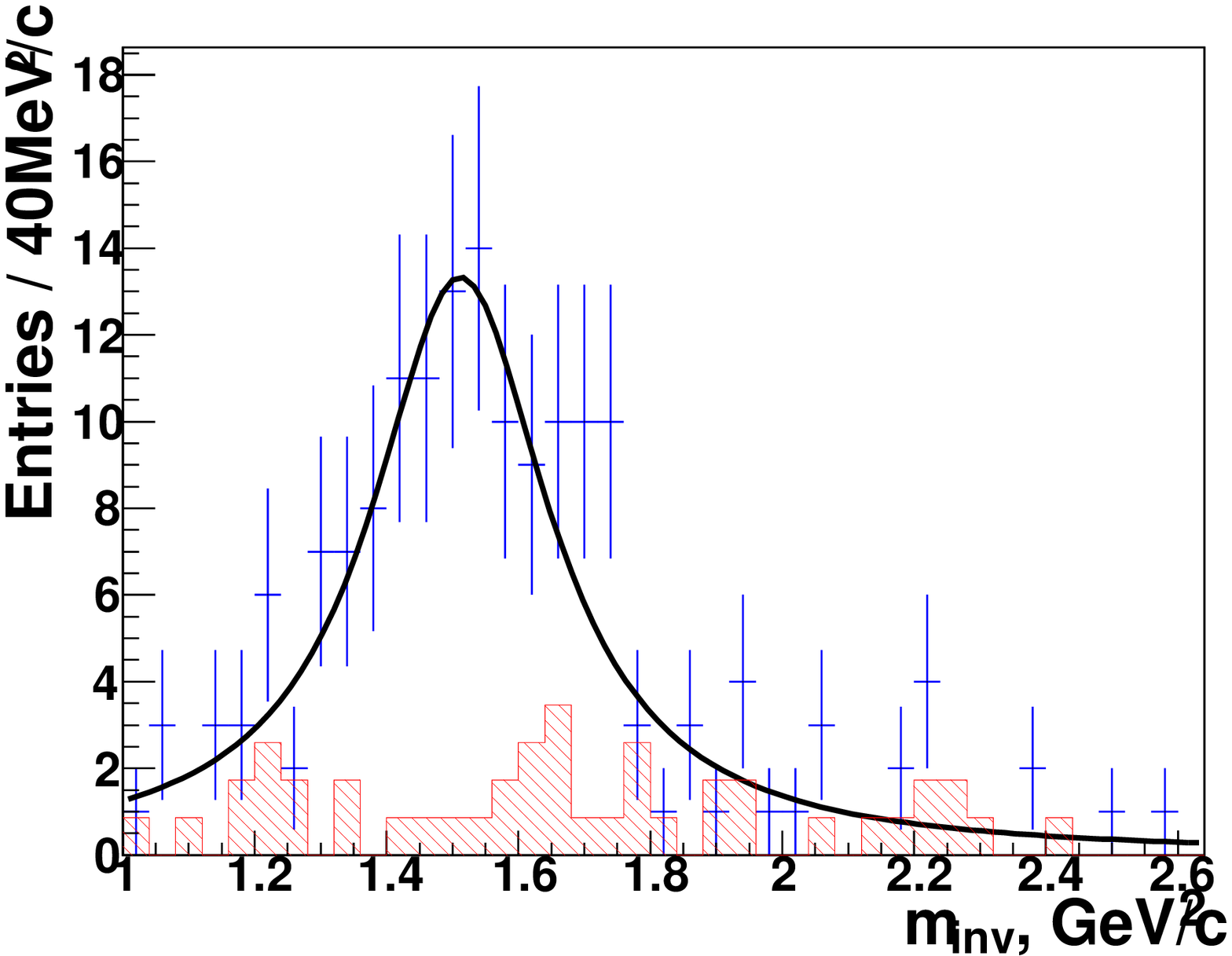}}
\put (30,60){\tiny STAR preliminary}
\put (130,65){\tiny STAR preliminary}
\normalsize
\end{picture}
\caption{The left panel shows the $p_T$ spectra of coherent 4-prong sample in the 2004 200 GeV Au-Au run.  Quads with total charge of zero are the gray histogram peaked at low $p_T$.  The hatched histogram shows the background estimated with charged quads.  For the invariant mass distribution, only those quads with total $p_T < 150$ MeV/c are selected.  There is a broad peak that may be consistent with a $\rho^{0*}$ meson like the $\rho^{0*}(1450)$. }\label{fig:rhoprime}
\end{figure}

We have analyzed $3.9 \times 10^6$ events from the 2004 200 GeV Au-Au run.  In this data sample we find on the order of 100 events with 4-track combinations that meet our criteria.  The $p_T$ distribution and invariant mass distribution are shown in Fig. \ref{fig:rhoprime}.  The $p_T$ distribution shows a peak at low total-$p_T$, indicating coherent production, and the invariant mass spectrum shows a broad peak around 1500 MeV, and may be consistent with a $\rho^{0*}$ resonance like the $\rho^{0*}(1450)$.

\section{Future Work}
	We expect to substantially increase the statistics for the 4-prong measurement in the 2007 200 GeV Au-Au dataset.  In addition, we hope the 2007 dataset will provide sufficient statistics to study rarer processes such as $J/\Psi$ production.

Roman Pots from the pp2pp experiment~\cite{pp2pp}  at RHIC are being moved to the STAR location in time for for the 2008 polarized proton run.  These will provide STAR with the capability to study diffraction in polarized pp collisions.

In the next few years, STAR will replace the Central Trigger Barrel with a Time-of-Flight detector.  This is expected to provide comparable triggering capabilities for ultra-peripheral collisions.  We also plan to upgrade our data acquisition system to permit Level-0 trigger rates of up to 1000 Hz, a substantial increase.

\section{Conclusions}
	STAR has measured both coherent and incoherent photoproduction of $\rho^0$ in AuAu collisions at 200 GeV.  We have compared the rapidity distribution as well as the total cross section with predictions from three different theoretical models.  While we cannot distinguish between the models on the basis of the measured rapidity distribution, we find that the measured total cross section favors the Klein-Nystrand model.  We have also measured the spin density matrix elements and found that they are consistent with s-channel helicity conservation.  Finally, we have observed $\gamma Au \rightarrow  \pi^+\pi^-\pi^+\pi^-$ production in AuAu at 200 GeV.

\section{Acknowledgments}
We thank the U. S. DoE-EPSCoR office for partial support of this work.

\begin{footnotesize}

\end{footnotesize}

\end{document}